\documentclass[traditabstract]{aa} 
\usepackage{txfonts,psfig}
\usepackage{color}

\begin{document}

\title{Richness-based masses of rich and famous galaxy 
clusters\thanks{Table 1 is available in electronic form
at the CDS via anonymous ftp to cdsarc.u-strasbg.fr (130.79.128.5)
or via http://cdsweb.u-strasbg.fr/cgi-bin/qcat?J/A+A/xxx}}
\titlerunning{} 
\author{S. Andreon\inst{1}}
\authorrunning{S. Andreon}
\institute{
INAF--Osservatorio Astronomico di Brera, via Brera 28, 20121, Milano, Italy,
\email{stefano.andreon@brera.inaf.it} 
}
\date{Accepted ... Received ...}
\abstract{
We present a catalog of galaxy cluster masses derived by exploiting
the tight correlation between mass and
richness, i.e., a properly computed number of bright cluster galaxies.  
The richness definition adopted in this work is properly calibrated, 
shows a small scatter with mass, and
has a known evolution, which means that we can
estimate accurate ($0.16$ dex) masses more precisely than by adopting any
other richness estimates or X-ray or SZ-based
proxies based on survey data. We measured a 
few hundred galaxy clusters at $0.05<z<0.22$
in the low-extinction part of the Sloan Digital 
Sky Survey footprint
that are in the 2015 catalog of Planck-detected
clusters, that have a known X-ray emission,  that are in the 
Abell catalog, or 
that are among the most most cited in the literature. 
Diagnostic plots and direct images of clusters are individually inspected
and we improved cluster centers and, when needed, we revised redshifts.
Whenever possible, we also checked 
for indications of contamination from other clusters on the line of sight,
and found ten such cases. All this information, with
the derived cluster mass values, are included in the distributed value-added 
cluster catalog of the 275 
clusters with a derived mass larger 
than $10^{14}$ M$_{\odot}$. 
A web front-end is available\thanks{At the URL
http://www.brera.mi.astro.it/$\sim$andreon/famous.html .}. Finally,
in a technical appendix we illustrate
with Planck clusters how to minimize the sensitivity of 
comparisons between masses listed
in different catalogs to the specific overlapping of
the considerd subsamples, a problem recognized but not solved in 
the literature. 
}

\maketitle

\section{Introduction}

Since Abell (1958) and Zwicky (1961) we have known that
the most massive clusters are also the richest.
A precise quantification of the relation between mass and richness
and, in particular, of the scatter between them has taken
a long time to be established because direct
estimates of mass not relying on hydrostatic or dynamical equilibrium,
such as caustic (Diaferio \& Geller 1997) or lensing 
(Broadhurst, Taylor, \& Peacock, 1995) masses, has
only recently become available and robust (e.g.,  Serra et al. 2011, 
von der Linden 2014a).

This quantification is still in progress for most richness 
estimates. For example, ``maxBCG" richnesses
(Koester et al. 2007) a) include fore/background galaxies 
among cluster galaxies (Koester et al. 2007; Andreon \& Hurn 2010); b)
count galaxies within a noisy radius on
average two times larger than $r_{200}$\footnote{The radius $r_\Delta$ is the
radius within which the enclosed average mass density is $\Delta$
times the critical density at the cluster redshift.}
(Becker et al. 2007; Johnston et al. 2007; Sheldon et al. 2009; 
Andreon \& Hurn 2010); 
c) use an incorrect 
center 30\% of the time (e.g., Johnston et al. 2007; Andreon \& Moretti 2011); 
and d) for these reasons
show redshift-dependent systematics
(Becker et al. 2007; Rykoff et al. 2008; Rozo et al. 2009; Andreon \& Hurn 2010;
Andreon \& Moretti 2011). Some other richnesses lack a mass
calibration (e.g., those in Budzynski et al. 2014), 
have a preliminary one (e.g., 
those in Rykoff et al. 2014),
have an unknown scatter with mass (e.g., those in Wiesner et al. 2015, Ford et al.
2015, and Oguri 2015),
or a yet unquantified evolution (e.g., those in Koester et al. 2007).
In these conditions, as for the ``maxBCG" richnesses mentioned
above or for richnesses in Rykoff et al. 
2014 (see Andreon 2015 for the latter), systematics may appear. 
Not fully calibrated richnesses are
not ready to be used to estimate cluster masses.

Other mass proxies share some of these shortcomings.
For example, the integrated
pressure $Y_X$ used by Vikhlinin et al. (2009) to estimate cluster masses
and, from these data cosmological parameters, has a yet uncharacterized evolution. 
In fact, Israel et al. (2014) found an higher normalization
interpreted as possibly due to a Malmquist bias. The current
calibration of the $Y_{SZ}$ mass proxy returns cosmological parameters different from
those derived mostly from the cosmic microwave background, which could
suggest a
calibration problem of up to 0.2 dex (Planck collaboration 2015),  or, using
independent data, a
mass bias (von der Linden 2014b) or a neglected evolution (Andreon 2014).
The mass calibration of the South Pole significance parameter largely
relies on simulations, given the scarcity of real data (Bocquet et al. 2015).

One proxy, $n_{200}$, seems to be in better shape -- it has a well-determined mean
scaling, a known and small scatter, and a known and negligible evolution -- as
a result of calibration efforts using
caustic masses from
the Cluster Infall Regions (Rines et al. 2006), Hectospec Cluster
Surveys (Rines et al. 2013), and
weak lensing masses from Canadian Cluster Comparison Project (Hoekstra et al. 2012).
A negligible scatter (0.02 dex) between a properly measured  
number of red galaxies and mass has been found 
around the relation
\begin{equation}
\log M_{200} = 14.86\pm0.03+(1.30\pm0.10) (\log n_{200} -2.0)
\end{equation}
with a tight upper limit on evolution with redshift up to $z=0.55$ once
the passive evolution of the red galaxies is accounted for 
(Andreon \& Hurn 2010, Andreon \& Congdon 2014, Andreon 2015). 
This $0.02$ dex intrinsic scatter in mass is comparable to or better 
than the values derived for other proxies, such as the integrated pressure $Y_{SZ}$ 
or pseudo pressure $Y_X$, X-ray luminosity $L_X$, gas mass $M_{gas}$, and 
stellar mass (Andreon 2015). The $n_{200}$ proxy performance, $0.16$ dex, is
comparable to or better than $Y_{SZ}$, maxBCG $n_{200}$ (Andreon 2015),
and $L_X$ (Andreon \& Hurn 2010). We note
that Eq.~1 refers to $n_{200}$ values measured according
to the prescriptions of Andreon \& Hurn (2010) and not to other
richness measurements even if they share the same symbol, such as 
``maxBCG" richnesses (Koester et al. 2007). 

A small scatter between proxy and mass makes the proxy
useful for predicting the mass of a cluster without a direct mass estimate.
However, given that the proxy has to be measured within an aperture which is nothing
else than a mass expressed in different units (for example, $r_{200}=M^{1/3}_{200}$
apart from obvious coefficients), an effective way to estimate
the reference aperture ($r_{200}$) is needed in order to use proxies to estimate
masses.  Andreon (2015) applies Kravtsov et al. (2006) idea of 
inferring both $M_{200}$ and $r_{200}$
at the same time exploiting the tight mass--proxy scaling. This
was shown to minimally
degrade the performance of richness as a way to estimate mass, leading
to richness-based masses with $0.16$ dex errors (Andreon 2015). 
The
observationally inexpensive richness can therefore be used to estimate the mass
of large samples of clusters for which either direct estimates are unavailable, 
are impossible to obtain, or just are not necessary. Unlike 
other mass proxies affected by dynamical or hydrostatic non-equilibrium,
the cluster richness (the number
of red galaxies) has the further advantage of also providing the mass of
clusters out of equilibrium (interacting, merging, etc.).

A large sample of clusters with homogeneously derived and calibrated masses
has many possible uses: 1) understanding how cluster properties
scale with mass; 2) performing other measurements of radial-dependent quantities
at a fixed reference radius, such as the fraction of blue galaxies
or the X-ray luminosity in a standardized aperture 
(as mentioned, $r_{200}$ can be simply derived from
the $M_{200}$); 3)
normalizing measurements whose definition requires knowledge of mass
such as the star formation density; 4)
combining clusters of different masses (and therefore sizes); 5)  
making available for study a larger sample of clusters with masses, allowing
the characterization of trends that are hard to identify with
smaller samples; 6) calibrating or checking the calibration of other
mass proxies; etc.

In this paper, we measure the mass of 275 clusters at $0.05<z<0.22$ 
in the Sloan Digital Sky Survey (hereafter
SDSS) footprint at high Galactic latitude. 
This measurement requires that we first improve the approximate
center and redshift of some of them. During the analysis, we also  
discover pairs of clusters on almost the same line of sight, which we list for
later use (some other mass proxies and direct mass estimates
are badly affected by this type of blends,  such as weak lensing
and SZ masses).
The resulting sample differs from 
many catalogs of clusters with known richness because
the reduced scatter of the 
adopted richness and the known calibration with mass allow us to
derive accurate ($0.16$ dex) masses. 

The appendix compares richness-based masses to SZ-based masses for 107
clusters. To perform this comparison, 
we solve 
the common, yet unsolved, problem of minimizing the sensitivity of conclusions
to the specific overlapping of the considered samples.

Throughout this paper, we assume $\Omega_M=0.3$, $\Omega_\Lambda=0.7$, 
and $H_0=70$ km s$^{-1}$ Mpc$^{-1}$. Magnitudes are in the AB system,
and all logarithms are in base 10.
We use the 2003 version of the Bruzual \& Charlot (2003) stellar 
population synthesis models with solar metallicity, a Salpeter initial 
mass function (IMF), and $z_f=3$.

\begin{figure}
\centerline{
\psfig{figure=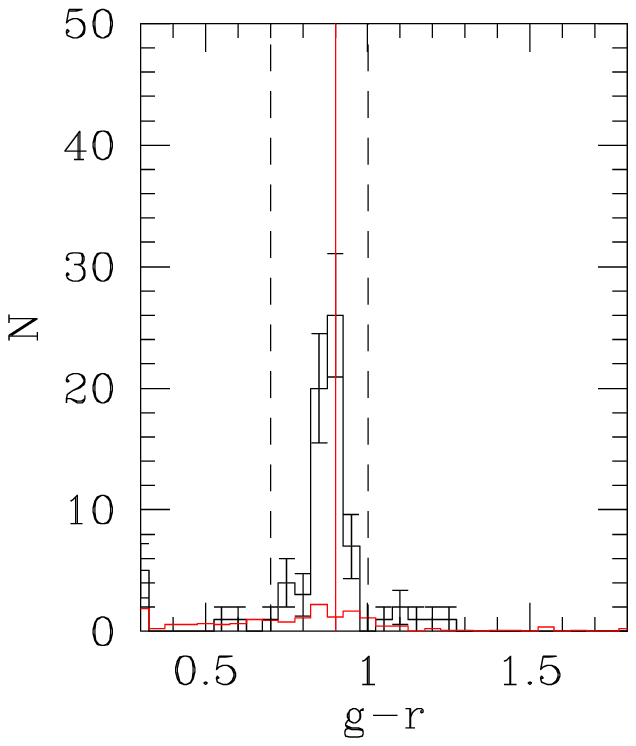,width=4.5truecm,clip=}
\psfig{figure=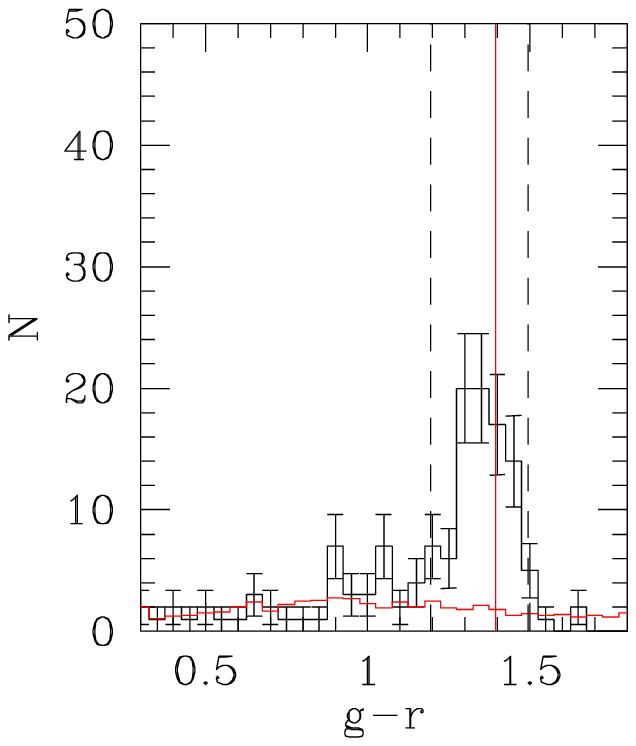,width=4.5truecm,clip=}
}
\caption[h]{Color histogram for the third nearest (left--hand panel)
and third most distant (right-hand panel) clusters. 
The black (red) histogram is the color distribution in the cluster
(control field) line of sight. The vertical red line
indicates the expected color of an old passively evolving population,
whereas the dashed lines mark the color
range where red galaxies are counted.  Error bars are  
$\sqrt{n}$-based for illustration only. Error bars of the control
field color distribution are omitted to avoid crowding, and are negligible
with respect to the cluster color distribution.
}
\end{figure}

\section{Cluster sample and derivation of cluster richness and mass} 

Our starting sample considers clusters satisfying the following four
conditions (logical ``and" operator):
\begin{enumerate}
\item
Thet are
listed a) in the Piffaretti et
al. (2011) compilation of X-ray detected clusters and with $\log L_X\gtrsim 43.5$ erg s$^{-1}$
in the 0.1--2.4 keV band, where the X-ray threshold value is set
to focus on massive clusters; 
b) in the Planck 2015 Catalog of Sunyaev-Zeldovich Sources (Planck
collaboration 2015); 
c) among clusters with at least 35 references in NED and $|b|>30$ deg;
and d) the Abell (1957) clusters with
a redshift in NED and  $|b|>30$ deg. 

\item
They are well inside the 
SDSS $12^{th}$ data release (Alam et al. 2015) footprint, i.e., 
with centers more than 1 deg away from the SDSS boundary, and not
severely masked by bright stars. 

\item
They have a redshift in the range $0.05<z<0.22$, the lower redshift boundary being set by the SDSS
shredding problem, the high redshift boundary by the SDSS depth.

\item
They have low Galactic extinction, defined as $A_r<0.5$ mag.
\end{enumerate}

For each of these clusters, we derive $n_{200}$ (richness) and, in turn, mass $M_{200}$
strictly following Andreon (2015) to which we refer for details. 
Basically, we count red members within a specified luminosity range
(only galaxies whose passive evolved magnitude 
is brighter than $M_{V,z=0}=-20 $ mag)
and color range (within $0.1$ redward and $0.2$ blueward in $g-r$ of the 
color--magnitude relation, our operational definition of ``red"), as already done for other
clusters (e.g., Andreon \& Hurn 2010; Andreon et al. 2014). 
For each cluster, we extracted the galaxy catalogs 
from the 
SDSS $12^{th}$ data release (Alam et al. 2015) and we
used ``cmodel" magnitudes for total galaxy magnitudes and 
``model" magnitude for colors. Colors are corrected for
the color--magnitude slope (but this is a minor correction 
given the small magnitude range explored). 
Fig.~1 shows, for the third 
nearest and most distant clusters, the color distribution and the adopted
color ranges. 

\begin{figure}
\centerline{
\psfig{figure=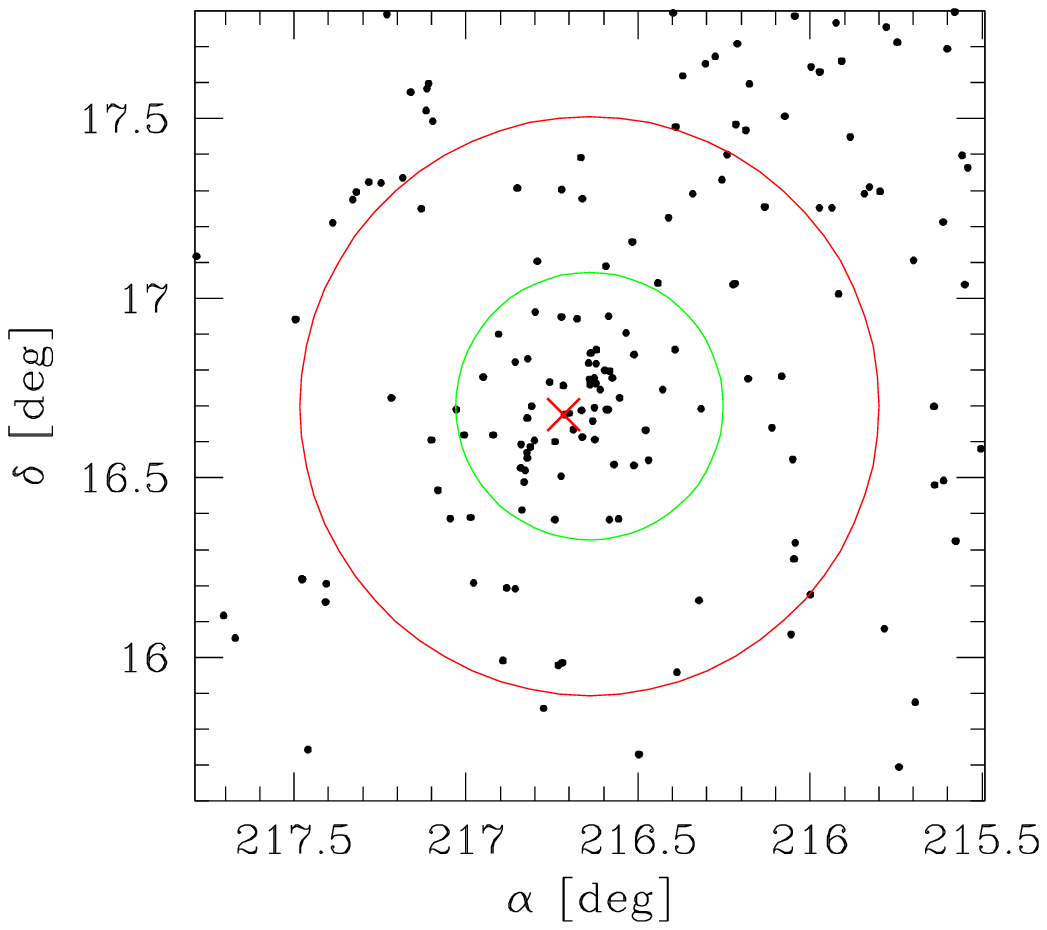,width=4.5truecm,clip=}
\psfig{figure=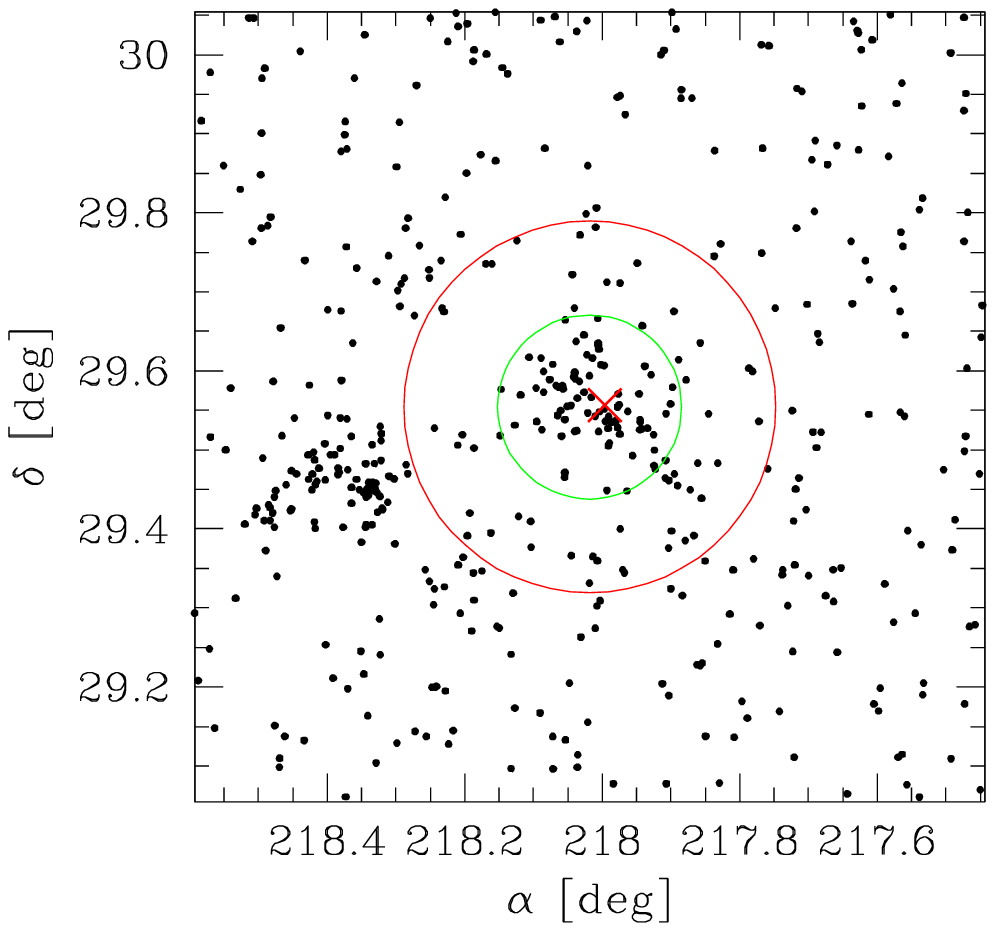,width=4.5truecm,clip=}
}
\caption[h]{Spatial distribution of red galaxies of the third nearest (left--hand panel)
and third most distant (right-hand panel) clusters. The inner (green) circle
marks the derived $r_{200}$ radius, the outer (red) circle indicates
3 Mpc at the cluster redshift. The cross indicates the cluster center as given
in the literature. In the right panel there is another obvious
galaxy overdensity, at $(\alpha,\delta)\approx (218.4,29.45)$.
}
\end{figure}

Some of the red galaxies in the cluster line of sight are actually in
the cluster fore/background. The contribution from background galaxies is
estimated, as usual, from a reference direction (e.g., Zwicky 1957; Oemler
1974; Andreon, Punzi \& Grado 2005). The reference direction is 
formed by three octants, free of contaminating structures (other clusters)
and not badly affected by the SDSS imaging masks, 
of a corona centered on the studied cluster with inner radius 3 Mpc and 
outer radius 1 (if $z>0.07$) or 2 (otherwise) degree(s), hence fully guaranteeing 
homogeneous data for cluster and control field. The color distribution of background 
galaxies, normalized to the cluster solid angle, is
shown in Fig.~1 for the two example clusters.

\begin{table*}
\caption{Clusters position, redshift, mass, and cross-identifications.}
\scriptsize
\begin{tabular}{l r r l l r r l l l l}
\hline
ID & R.A. & Dec. & z & $\log M/M_\odot$ & $n_{200}$ & err & Piffaretti et al. 2011 ID & Planck 2015 ID & Abell 1957 ID & NED 2015 ID  \\
& \multicolumn{2}{c}{J2000} & & &  \\
\hline
GCwM 1 & 0.8380 & 4.6440 & 0.098 & 14.05 & 24 & 5 &			      & 			  &	    ABELL2698 (1.9)   & ABELL2698 (1.9,17)			\\ 
GCwM 2 & 0.9235 & 2.0678 & 0.092 & 14.26 & 34 & 6 &  MCXCJ0003.8+0203 (2.3)   & PSZ2G099.57-58.64 (1.5)   &	    ABELL2700 (2.6)   & ABELL2700 (2.3,66)		       \\
GCwM 3 & 1.3754 & 16.2435 & 0.116 & 14.29 & 36 & 7 &  MCXCJ0005.3+1612 (2.7)   &			   &			       & WHLJ000524.0+161309 (2.1,21)	       \\ 
GCwM 4 & 1.5851 & 10.8613 & 0.170 & 14.49 & 52 & 8 &  MCXCJ0006.3+1052 (0.5)   & PSZ2G105.40-50.43 (2.0)   &			       & NSCSJ000619+105206 (0.5,10)	       \\ 
GCwM 5 & 2.5753 & 17.7496 & 0.173 & 14.47 & 51 & 8 &			       & PSZ2G109.22-44.01 (1.2)   &		ABELL6 (4.2)   &				       \\ 
GCwM 6 & 2.9235 & 32.4316 & 0.107 & 14.61 & 64 & 9 &  MCXCJ0011.7+3225 (1.1)   & PSZ2G113.29-29.69 (0.8)   &		ABELL7 (1.2)   &				       \\ 
\hline      																					    
\end{tabular}
\hfill\break
The table lists cluster coordinates (J2000), 
redshift $z$, $\log$ mass $M_{200}$ estimated from the richness,  the richness $n_{200}$ and its error, 
followed by the id of clusters
from a few bibliographic sources, with angular offsets reported in parentheses. The
last column also lists the number of references to the cluster in Mai 2015, as given by NED. 
Masses have $0.16$ dex errors.
The table is entirely available in electronic form
at the CDS, 
and also accessible at http://www.brera.mi.astro.it/$\sim$andreon/famous.html 
. A portion is shown
here for guidance regarding its form and content.																					    
\end{table*}

The centers given in the literature are sometimes imprecise: Abell
(1958) estimated them by eye, Planck has a poor point spread function
and therefore large (1.5 arcmin) positional errors, and Piffaretti et al.
and NED collect positions derived with a variety of methods from
various sources sometimes having poor resolution or affected
by unrelated point sources (as the Rosat All
Sky Survey is).
We estimate the cluster center iteratively as the median values of 
right ascension and declination 
of red galaxies within an aperture of 1.0 Mpc radius, starting the iteration
on the literature center. We then iterate 11 times and take
the last value as the final center. The initial and final center are
indicated in Fig.~2 for the two example clusters.
Andreon (2015) showed that results do not change when using another number of 
iterations because convergence is achieved earlier, whereas
in Sec.~4 we show that the starting position does not matter when
using duplicate clusters (objects with center offsets larger than 3 arcmin
in different cluster catalogs, and therefore listed
twice in our initial list of clusters to be analyzed). We
note, however, that X-ray, SZ, and other types of optical centers may legitimately 
be different from our centers based on galaxy numbers.

As mentioned, the $r_{200}$ radius is unknown for the studied clusters.
We adopt, as proposed in Andreon (2015) for richness and in Kravtsov et al. (2006)
for $Y_X$, an iterative approach to its determination,
which exploits the almost scatter-less
nature of the richness--mass relation: a radius $r$ is
taken (1.4 Mpc in our case), $n(<r)$ estimated, then $r$ is updated
to the value appropriate for the derived richness (i.e., using eq. 1, 
and noting that $r_{200}=M^{1/3}_{200}$
apart from obvious coefficients) and then the process is iterated $3$ times.
This procedure returns $M_{200}$ with a total scatter of only 0.16 dex
from true (Andreon 2015).
The derived $r_{200}$ is shown (inner circle) for the two example clusters in Fig.~2.
In Andreon (2015) we show that adopting a different number of iterations
does not change the results because convergence occurs earlier.

Since the mass-redshift relation is known to hold for $\log M_{200} /M_{\odot}> 14$
(and we ignore if it holds at lower masses),
only clusters more massive than this threshold are kept in the sample. 
To be more precise, the applied cut is $\log n_{200}>21.6$.
This leave us with a sample of 275 clusters.
Figure 3 and 4 show the distribution of these 275 clusters 
in the sky and in the redshift--mass 
plane.

\begin{figure}
\centerline{
\psfig{figure=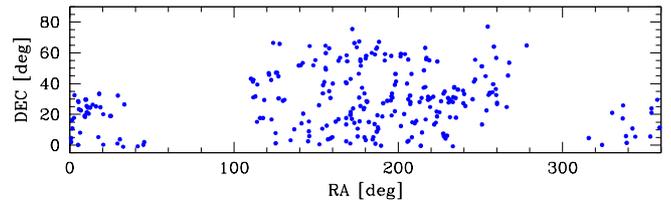,width=9truecm,clip=}
}
\caption[h]{Sky distribution of the studied cluster sample. The SDSS footprint
is clearly imprinted.
}
\end{figure}

The values of masses derived in this way are listed in Table 1 (entirely available in
electronic form at the CDS and also at
http://www.brera.mi.astro.it/$\sim$andreon/famous.html with a
front-end to the SDSS imaging service). These masses are the prime result
of this work: while many literature
papers report richnesses for cluster samples, these
richnesses have a larger scatter with mass, often with an unknown
or problematic mass calibration.
The current work instead uses a precisely calibrated richness--mass
scaling using a small scatter proxy. 
Radii can be derived using
the $M_{200}$ definition,
whereas the originally measured richnesses are also listed. 
We remember that richness is within a cylinder of radius $r_{200}$, whereas
mass is deprojected (i.e., within a sphere) because the calibration,
i.e., Eq.~1, returns the latter as a function of the former.
Table~1 also lists other known 
identifications of the studied clusters when their reported
coordinates are within 3.0 arcmin from the center determined in this work (5.0
armin for Abell 1957 and NED). 
The angular offset, in arcmin, is reported in parentheses.
Table 2 shows the number of clusters
in each subsample (Piffaretti et al. 2011, Planck 2015, Abell 1957 and NED)
and in their overlaps.
Comments to a few individual clusters are listed in Appendix A.

Equation~1, used to estimate mass from richness, has a zero evolutionary term,
in agreement with the small and statistically insignificant 
term determined in Andreon (2015), $-0.1\pm1.0$, and with the 
tighter constraint derived in Andreon \& Congdon (2014). A ten times larger term, $1.0$,
would induces a systematic error of $\pm0.03$ dex across the studied redshift range,
negligible compared to our quoted error $0.16$ dex.
For this reason, we neglected the $-0.1\pm1.0$ evolutionary term found 
in Andreon (2015).

\begin{figure}
\centerline{
\psfig{figure=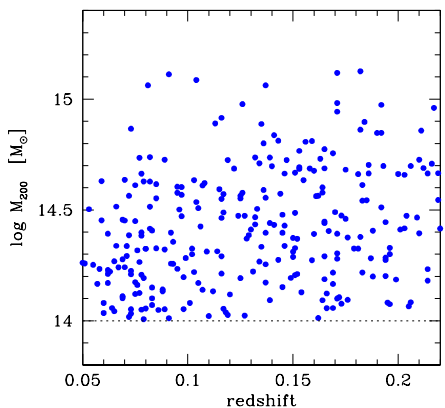,width=6truecm,clip=}
}
\caption[h]{Mass vs redshift plot of the studied cluster sample. 
Mass has 0.16 dex errors, omitted to avoid crowding. 
}
\end{figure}

\begin{table}
\caption{Number of clusters in each subsample}
\begin{tabular}{l r r r r }
\hline
  & Piffaretti et al. & Planck  & Abell  & NED    \\
\hline
Planck  &        72   &	    107  &	   89   & 	 90 \\
Abell   &      119    &	    89   &	 211    &      197 \\
NED     &     141     &	   90	 &	197     &     246 \\
\hline
Total 	     &         155 & 	     107& 	   211 & 	 246 \\
\hline      																					    
\end{tabular} 																					    
\end{table}																					    

\subsection{Can we apply the calibrating mass-richness relation to
different cluster samples?}

The mass calibration in Andreon (2015),
i.e., Eq.~1, uses the observed number of galaxies to predict the mass of 
the cluster, and has been derived for a calibrating sample with a given
selection function. One may therefore wonder if this may be used
for the current, target, sample with a different selection function. The
answer is yes, as we now illustrate.
We consider
the general case of a proxy $x$ whose observed value has been
calibrated against mass $M$
\begin{equation}
\log M  \propto \log x^{obs} 
\end{equation}
with some scatter $\sigma_x$, where we have absorbed in $x$ the intercept
and the possibly non-zero slope as well as the effect of the selection function and
steep mass function of the {\it calibrating} sample. 
This relation does not change when
another variable $y^{obs}$ showing some scatter $\sigma_y$
with $\log M$
\begin{equation}
\log M  \propto \log y^{obs} 
\end{equation}
is considered, available, and satisfies some
constraints, which means that the cluster is included in the {\it target} 
sample. In fact, for
each individual cluster, the relation in Eq.~2 is unaltered,
neither the value of $x^{obs}$ nor $M$ changes because another cluster
property $y^{obs}$ is measured. The quantity $x^{obs}$ still provides an unbiased
estimate of the cluster mass via Eq.~2. 
This make us free to
use Eq.~2 (and Eq.~1, which has the coefficients explicitely given)
for a sample selected by another observable\footnote{We missed this point
in Andreon (2015), and therefore we were overly restrictive in that paper
about the applicability of richness as mass proxy.}. 

However,
the data cloud satisfying the selection on $y^{obs}$ can obey a different
relation as a result of sample selection effects. This can be easily understood
when only $\log M>15$ clusters are kept in the sample, and
where mass is estimated from a proxy $y$ with negligible scatter with mass.  
At low $x^{obs}$ the $M-x^{obs}$ relation would no longer decrease with 
$x^{obs}$, but instead would flatten at $\log M\sim15$
because clusters with $\log M<15$
are excluded by the selection (and $\log M\gg 15$ with low $x^{obs}$ become 
exceptionally rare). 
Eq.~2 still provides
an unbiased mass, but because of important selection effects the proxy vs mass
trend of the selected sample
will appear to differ from a linear relation with slope one and zero offset,
as detailed for the Planck subsample in Appendix B.

We also emphasize that while richness, and therefore richness-based
masses, does not depend on the cluster status,
other mass proxies assume a hydrostatic or dynamical equilibrium and are 
affected by an unknown hydrostatic bias. This too may lead to a possible systematic
offset, or scatter, with our richness-based masses.

\section{Value-added features of the cluster catalog}

When deriving cluster masses we performed an extensive set of quality
controls. Basically, every plot shown in this paper has been inspected
for each and every cluster; SDSS images were inspected as well as individual spectra
of the most important galaxies (typically the brightest cluster galaxy). During these
controls we noted the following:
\begin{enumerate}

\begin{figure}
\centerline{
\psfig{figure=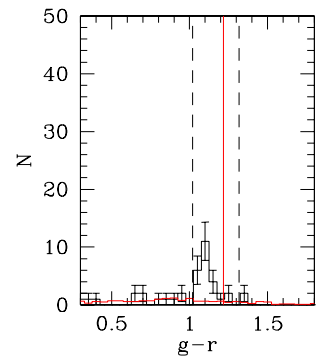,width=4.5truecm,clip=}
}
\caption[h]{Color histogram of Abell 1182. Symbols and error bars 
are as in Fig.~1,
but in this figure we adopted the literature redshift $z=0.166$ instead
of the SDSS spectroscopic-based $z=0.148$ we derived, showing
how the color of the red sequence may pinpoint an approximate literature redshift.
}
\end{figure}

\begin{figure}
\centerline{
\psfig{figure=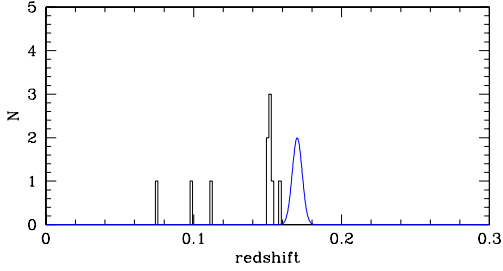,width=7truecm,clip=}
}
\caption[h]{Redshift distribution of galaxies projected within the
$r_{200}$ radius of GCwM 170 (alias PSZ2G114.83+57.25, MCXCJ1325.8+5919, Abell 1744). 
The blue Gaussian is centered on the 
cluster redshift as given in the literature. The brighter cluster galaxy and five more galaxies
are instead at $\Delta z =-0.02$ from the reported value.  
}
\end{figure}

\item
The cluster redshift listed in the literature (which comes from a variety
of sources, including photometric redshift and perhaps sometimes from the
redshift of one
single galaxy) is inaccurate for 21 clusters
by an amount that is easy to note for at least one
of two reasons. First, the
observed red sequence is not at the expected color. This is
illustrated in Fig.~5 for Abell 1182: the observed red sequence is bluer
than it should be for a $z=0.166$ cluster. Indeed, the cluster is at $z=0.148$,  
based on SDSS spectroscopy, a $\Delta z\sim0.02$ offset from the literature 
(photometric in this case) redshift. If not corrected
for, such a small offset might lead to an overestimation of the cluster 
mass of 0.09 dex, smaller than our quoted mass error. We 
note that on some rare occasions redshift offsets as small as $\Delta z=0.004$
(as for Abell 1045, also known as MCXCJ1034.9+3041) have been detected from
the color offset. Second, we
queried the spectroscopic SDSS database and checked whether the 
redshift peak of the galaxies within $r_{200}$
lies at the literature redshift. If not, we updated the literature redshift 
(an example is shown in Fig.~6, also in this case the redshift change
lead to a negligible mass change). 
Of course, cluster parameters in Table 1 are estimated with
the revised redshift, although this is a negligible correction\footnote{Abell 1182
has a mass below the limit for inclusion in the final catalog, and therefore
is not listed in Table~1.}.

\item
Ten clusters listed in the literature are, on the $r_{200}$ 
spatial scale, 
blends of widely separated ($\Delta v > 1500$ km/s) clusters on
the same, or nearby, line(s) of sight. 
These are recognized for having
two red sequences at different colors and, when sufficient
spectroscopic data are available, two or more redshift peaks (see Fig.~7 for
an example). The ten blends are listed in Table C.1. This list constitutes
a further result of this work because
these clusters are likely blended in SZ too given the 
large Planck PSF and the lack of redshift sensitivity, as well as
in weak--lensing analysis (again because of the poor redshift
sensitivity of shear). Indeed, one of this pair (Abell750/MS0906) is
a known case of lensing blending (Geller et al. 2013).
Seven out of ten have been 
detected via their intracluster medium in emission. Therefore,
it does not seem that X-ray selected are immune to 
projection effects 
as one often reads in literature papers. It is true and obvious, however, that 
confusion effects are lower when a
smaller aperture is adopted, as is compulsory in X-ray because of the shallowness of the
signal at large radii and as is also feasible for richness. In fact,
the clusters listed in Table C.1 could
be not contaminated on smaller spatial scales, such as
those probed by the X-ray emission and also by richness data
when adopting a smaller aperture. 

\begin{figure}
\centerline{
\psfig{figure=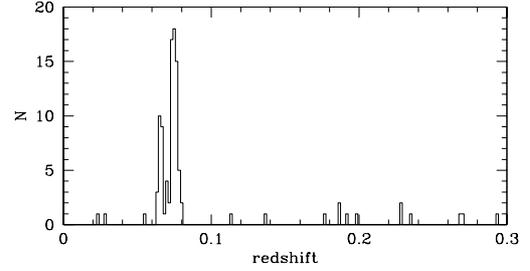,width=7truecm,clip=}
}
\caption[h]{Redshift distribution of galaxies projected within 
1.4 Mpc of PSZ2G031.93+78.71. The two redshift peaks are separated
by 2400 km s$^{-1}$.
}
\end{figure}

\begin{figure}
\centerline{
\psfig{figure=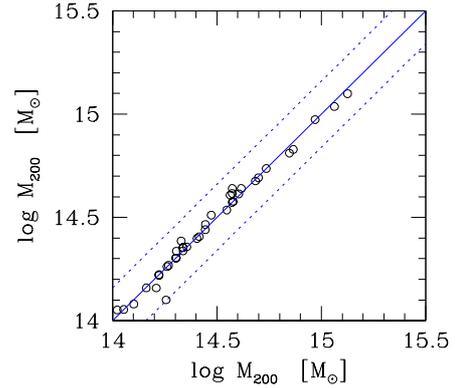,width=6truecm,clip=}
}
\caption[h]{Comparison of masses for 38 duplicate clusters (40 comparisons)
with center coordinates that differ by more than 3 arcmin in different
cluster catalogs. The solid line indicates equality, the corridor marks 
our mass error ($\pm0.16$ dex). 
}
\end{figure}

\item
On three occasions the SDSS photometry is corrupted, probably because
of a background subtraction problem. This shows up
as a rectangular region in the sky where galaxies are missing or have
inappropriate colors for their spectroscopic redshift, notably
a red sequence with the wrong color for the SDSS-measured spectroscopic
redshift. We found the
following clusters affected by the above, and therefore dropped from our list:
Abell 1682 (alias MCXCJ1306.9+4633, alias PSZ2G114.99+70.36), 
Abell 2029 (also noted in Renzini \& Andreon 2014), and
MCXCJ0751.4+1730 (alias Abell 598).

\end{enumerate}

\section{Checks}

A number of checks on the calibration and scatter of the richness-mass
relation was done in our previous work using sample with well-measured
(caustics, hydrostatic, or weak--lensing) masses. In addition to these,
\begin{itemize}
\item
38 cluster of our initial list of clusters to be analyzed (and with a 
derived mass larger than $\log M /M_{\odot}> 14$) were analyzed twice
because they have center coordinates that differ 
by more than 3 arcmin (175 to 640 kpc, depending on redshift) in
different cluster catalogs. These duplicate 
clusters (listed only once in
our final catalog) are useful for estimating the noise introduced by 
our mass estimate procedure because they use
different background areas and different starting centers. 
Figure~8 compares mass estimates for these duplicate
clusters (there are 40 because two clusters appear in three catalogs).
We found a scatter
between the different mass estimates of the same cluster 
almost 10 times smaller than
our quoted mass uncertainty. Therefore, the centering procedure and
background estimates introduce a negligible error, independently
confirming the tests in Andreon (2015). 

\begin{figure}
\centerline{
\psfig{figure=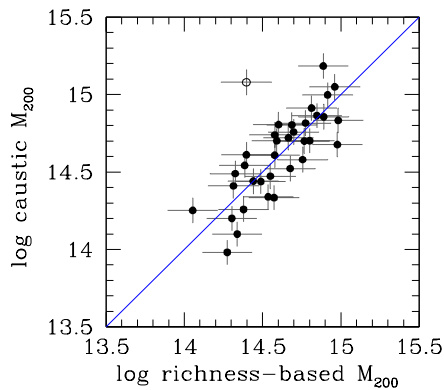,width=6truecm,clip=}
}
\caption[h]{Richness-based mass vs caustic mass. The former only uses SDSS photometry,
the latter abundant follow-up spectroscopy. The open point is Abell 1068, a cluster
with a caustic mass of dubious quality.  
}
\end{figure}

\item
35 clusters in Table 1 are also in the Hectospec Cluster
Surveys (Rines et al. 2013) sample used by
Andreon (2015) to calibrate the relation between mass and richness
(Eq.~1).
The predicted mass, derived in this paper, vs the caustic mass listed
in Rines et al. (2013) is shown in Fig.~9. The scatter  in mass at a
given richness is
$0.16$ dex, i.e., what we reported in a similar analysis based on almost the same
cluster sample. The outlier point is Abell 1068, having
a caustic mass in disagreement with past mass measurements (see
Andreon 2015 for details). This comparison
differs from our past one by the 
use of a different starting center and, most of the times,
a different background area. This comparison shows again that
the centering procedure and
background estimates introduce a negligible error into our masses. Comparison with
caustic masses in 
the Cluster Infall Regions (Rines et al. 2006) shows a larger
scatter because of the known lower precision of the centers used there
(Andreon \& Hurn 2010; Andreon 2015).

\item
Out of the 71 clusters with $Y_X$-based mass listed in 
Planck Collaboration et al. (2014), 11 are also in Table 1.
The mass comparison is shown in
Fig.~10. There is a good agreement, except for Abell 1795, a cluster with
a complex X-ray morphology (a cavity, a cold front, and a cooling wake; see
Walker et al. 2014 and Ehlert et al. 2015 and references therein).
Although the agreement is promising, we emphasize once more 
that because the parent sample (the 71 clusters with $Y_X$) does not have
a selection function, this comparison should be taken with caution, 
as discussed in detail for the larger sample of the Planck clusters in our
Appendix B.

\end{itemize}

\section{Summary and conclusions}

In this paper we exploit the tight correlation between richness and mass, calibrated
on more than one hundred clusters in Andreon (2015).
By simply counting the number of red galaxies brighter than the appropriate limit,
accounting for the fore/background, we estimated $M_{200}$ and $r_{200}$ of a
sample of 275 clusters with $\log n_{200}>21.6$ corresponding to
$\log M /M_{\odot}> 14$ in the low-extinction
part of the SDSS footprint and with $0.05<z<0.22$. 
Position, redshift, and more importantly
mass with a $0.16$ dex precision is given in Table 1 for the 275 clusters.
By adopting a low-scatter well-calibrated mass proxy,
this catalog
delivers masses whose precision exceeds those available by adopting any of the
many proxies, including other richnesses, available in the literature. 
Our mass measurements are
homogeneously derived using homogeneous data, making our catalog different from
literature collections of heterogeneous measurements derived with a variety
of methods using data of variable quality, such as the Piffaretti 
et al. (2011) or the Sereno (2015)
catalogs.

\begin{figure}
\centerline{
\psfig{figure=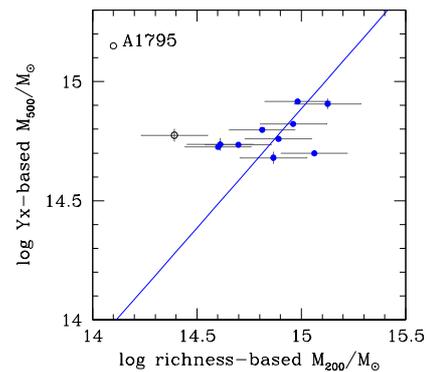,width=6truecm,clip=}
}
\caption[h]{Richness-based mass vs $Y_X$--based mass. The former only uses SDSS photometry,
the latter X-ray deep follow-up photometry and spectroscopy. The open point is Abell 1795, 
a well--known cluster with complex X-ray morphology.  
}
\end{figure}

The derived masses are useful for a variety of purposes: to understand how a cluster property
scales with mass, to perform measurements of radial-dependent quantities
at a fixed, reference, radius ($r_{200}$ can be simply derived from
the listed $M_{200}$ values),
to normalize measurements whose definition requires knowledge of mass,
to combine clusters of different masses, etc.
The derived richness-based masses are unaffected by deviation of the 
cluster from dynamical or hydrostatic equilibrium because the proxy, i.e., 
the number of red galaxies, is unaffected by these deviations.

During the analysis, a much larger sample of clusters
was individually scrutinized: about half of the analyzed 
clusters turned out to have $\log M /M_{\odot}< 14$, several clusters
were dropped because severely hidden by bright stars,
a revision was needed for about a quarter of the sample (clusters
with inaccurate redshifts or widely different centers in different catalogs),
and we identify 10 ``clusters" that are actually blends, on the $r_{200}$ angular
scale, of clusters at different redshifts on almost the same line of sight. 
These clusters are listed in Table C.1 for later use (other mass proxies are
similarly badly affected). 

Finally, in Appendix B we compare richness-based to SZ-based masses.
To achieve this purpose, we solve
the common, and yet unsolved, problem of minimizing the sensitivity of conclusions
to the specific overlapping of the considered samples.

\begin{acknowledgements}
I thank the referee for the comments emphasizing the strengths
of the catalog.
Lucia Ballo is thanked for proposing the acronym adopted to name the clusters.
This research has made use of the NASA/IPAC Extragalactic Database (NED).
\end{acknowledgements}

{}

\appendix

\section{Comments to individual clusters}

\begin{itemize}
\item{GCwM 14 (Abell 67):} possible multiple system
\item{GCwM 23 (Abell 115, MCXCJ0055.9+2622, PSZ2G124.20-36.48):} possible underestimated mass because of two bright stars
\item{GCwM 62 (Abell 665, MCXCJ0830.9+6551, PSZ2G149.75+34.68):} bimodal cluster
\item{GCwM 92 (Abell 1033, MCXCJ1031.7+3502, PSZ2G189.31+59.24):} bimodal cluster
\item{GCwM 116 (Abell 1307, MCXCJ1132.8+1428, PSZ2G243.64+67.74):} bimodal cluster
\item{GCwM 181 (Abell 1800):} bimodal cluster
\item{GCwM 188 (Abell 1882) and GCwM 189 (MCXCJ1415.2-0030):} these two clusters
have the same redshift and have centers that are separated by about 
the estimated $r_{200}$. If they are one single cluster, its mass is
larger than both quoted values: if they are two separate clusters the quoted masses
are overestimated. 
\item{GCwM 27, 99, 140, and 225} do not have any known identification in
the input catalogs as a result of extreme recentering offsets.
\end{itemize}

\section{Selection function, error, scatter, and mass function 
are important in mass--mass comparisons}

\begin{figure*}
\centerline{\psfig{figure=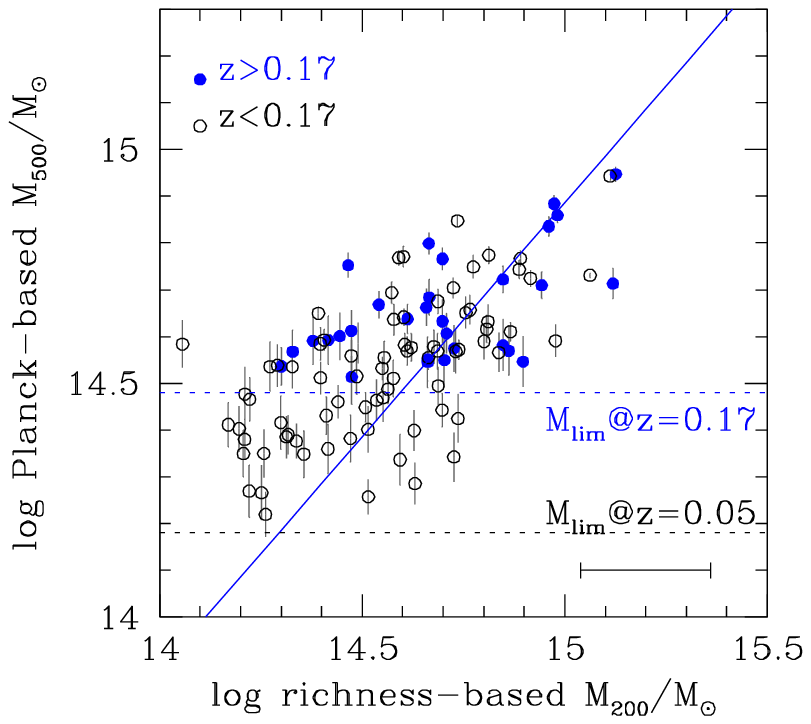,height=5.5truecm,clip=}%
\psfig{figure=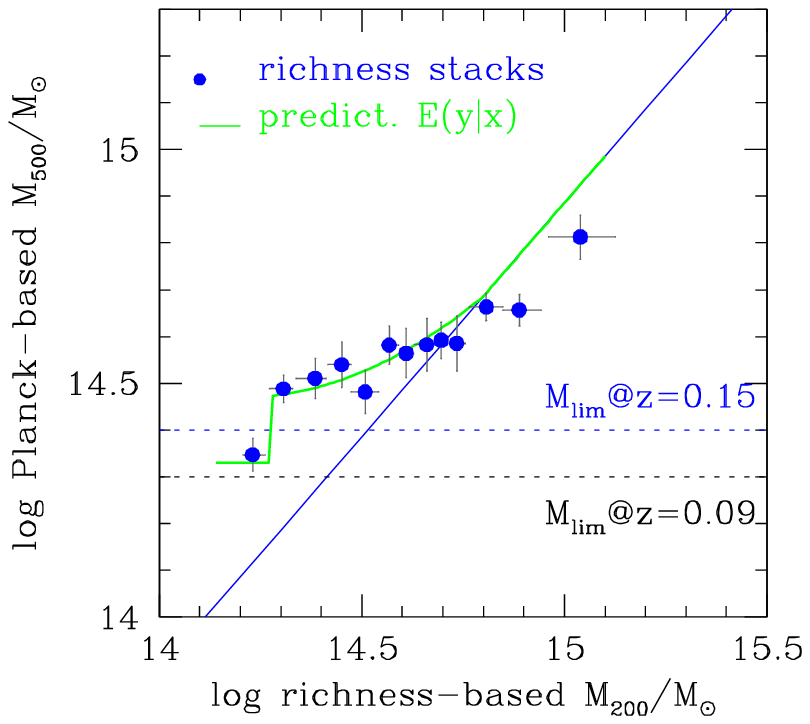,height=5.5truecm,clip=}%
\psfig{figure=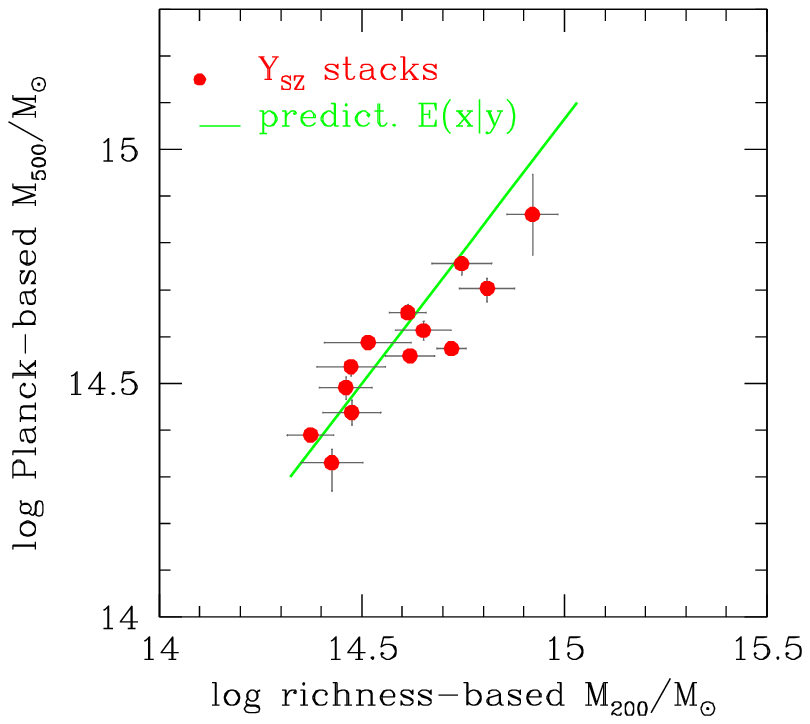,height=5.5truecm,clip=}}
\caption[h]{Richness-based $M_{200}$ vs Planck--based $M_{500}$.
{\it Left--hand panel:} 
Points are color- and symbol-coded according to redshift. 
To avoid crowding, the error on abscissa is indicated in the
bottom-right corner. 
{\it Central and Left--hand panels:} Clusters are coadded in bins of 8.
Bins are on the abscissa in the central panel, on the ordinate in the right--hand
panel. 
Errors on the binned axis indicate the bin size, errors on the other
axis is the error on the average
(the scatter divided by $\sqrt{8-1}$). The green line indicates the
expected trend based on simulations. 
In the left--hand and central panels,
the Planck limiting masses for two redshifts and the
relation $M_{200}=1.3 M_{500}$ 
are also
shown. 
}
\end{figure*}

It is just beginning to be recognized that the comparison between masses listed
in different catalogs may be sensitive to the specific overlapping of
the considered subsamples (e.g., Okabe \& Smith 2015). Ignoring  
the selection function of the overlap 
may lead to claiming the existence of systematic differences between masses measured
by different authors/methods/proxies when differences are instead
the effect of the selection induced by only considering the overlap between
the compared samples. Or, one may also miss an obvious systematic
difference cancelled out by a selection effect. 
In this appendix we go beyond the common 
and generic statements above  
by showing how to account for
the selection
using the subsample of Planck clusters. We emphasize that a naive
comparison, not accounting for selection effects, would have claimed the masses
derived here to show systematic differences with those derived by Planck
whereas the found behavior is instead a manifestation of selection effects
induced by Planck only seeing very massive clusters.

Our analysis makes some simplistic assumptions.
In particular, we assume a) uncorrelated error and scatter between richness
and $Y_{SZ}$; b) a step function to describe  
the Planck selection function; and c) that the Planck selection
function does not depend on other other astronomical
observables (such as Planck exposure time). This suffices to illustrate 
the dangers of a naive comparison: the observed trend may differ 
from a one-to-one relation without tilt
or bias.

One handred and seven clusters in our sample have a match ($<3'$ offset) with a cluster
in the second Planck catalog of SZ sources (Planck Collaboration 2015a),
61 of which are in the cosmological sample. These clusters are listed
in Table 1. These are a random sampling (in mass) of the Planck catalog because
we analyzed all Planck clusters inside the SDSS (with $0.05<z\lesssim0.22$, depending
on Galactic extinction) without retaining or discarding any of them
because of their mass. In fact, our only mass-dependent selection is
$\log M_{200} /M_{\odot}>  14$, which is far more liberal than
the original Planck catalog, $\log M_{500} /M_{\odot}\gtrsim 14.4$ (we also note the
higher overdensity used for Planck masses). This can also be
seen in Fig.~B.1: there are no clusters at $\log M_{200} /M_{\odot}\sim 14$
because they are below the Planck sensitivity,
while many of them are in our sample (see Fig.~4).
The studied clusters are precisely 
in the redshift range where Planck Collaboration (2015b) found a deficit
of observed cluster counts for the cosmology (mostly) inferred from
the cosmic microwave background (and therefore of utmost interest). 

The comparison of the two mass estimates of
individual clusters is shown in the left-hand panel of Fig.~B.1. 
The two estimates
of mass are correlated, although with some scatter, indicating that both
are tracing the cluster mass. The trend identified by the points deviates
by the naive expectation, a trend with slope one and intercept zero (depicted
as a solid line in the figure) after accounting for differences in the density 
$\Delta$ adopted for the compared masses.

However, the trend (slope, shape, and scatter
around the mean) are largely driven by the Planck selection effects
(which dominates over our sample selection, as mentioned). This can be 
easily guessed by
splitting the sample into two halves (below and above $z=0.17$), and noting
that points tend to ``turn left" of the $M_{200}=1.3 M_{500}$ (slanted) line
when approacing the relevant Planck limiting mass. These are 
the expected (and observed
in the case of simulated Euclid weak lensing masses, see Andreon \& Berg\'e 2012)
behavior of a sample affected by selection effects, as we now illustrate in detail.

A well--posed comparison of masses must account
for the large population gradient (i.e., the cluster mass function),
the unavoidable error, or scatter, of the mass proxies
(which induces
a Malmquist--like bias when joined to the population gradient, 
Andreon \& Hurn 2010; Andreon \& Hurn 2013; 
Andreon \& Weaver 2015), and the
Planck selection function. In passing, these make
the locus of the expected $Y_{SZ}$-mass at a given richness-mass, 
$E(\log M_{Planck}|\log M_{richness})$, different
from the locus obtained the other way around, $E(\log M_{richness}|\log M_{Planck})$.
Furthermore, one of these locii is not even a straight line for input straight 
relations, as we now show.

The expected $Y_{SZ}$-based mass vs richness-mass trends has been computed 
using a simulation, similar to Andreon \& Berg\'e (2012)
and Andreon \& Congdon (2015): we extracted masses from a Jenkins et al. (2001)
mass function at the mean redshift of the clusters entering in each mass bin
using the halo mass calculator (Murray et al. 2013).
About richness-based masses, their scatter is $0.16$ dex and therefore are scattered
by such an amount. Real {\it observed} data are calibrated on  caustic masses 
to have slope one and zero intercept. Therefore, we also calibrate simulated observed data
on true masses  (we do not simulate a caustic estimate)
in the same way. $Y_{SZ}$-based masses are derived from true
$M_{200}$ masses with $M_{200}=1.3 M_{500}$, 
i.e., what one expects for a Navarro, Frenk \& White (1997)
profile of concentration of about $5$ and
scattered by $0.06$ dex (the scatter of $Y_{SZ}$-based masses, Planck collaboration 2014), 
and then removed from the simulated sample if less massive than
the Planck limiting mass at the redshift of interest (e.g., $\log M_{500} /M_{\odot}= 14.4$
at $z\sim 0.15$, from 
Planck collaboration 2015a). 
With these simulated masses, we computed the mean $Y_{SZ}$-based mass in small bins
of richness-based masses (green line in the central panel of Fig.~B.1), 
$E(\log M_{Planck}|\log M_{richness})$, and
the other way around: the mean richness-based mass in small bins of $Y_{SZ}$-based mass,
$E(\log M_{richness}|\log M_{Planck})$ (green line in the right--hand panel of Fig.~B.1).
At low masses, because of the 
Planck selection function, $E(\log M_{Planck}|\log M_{richness})$ 
must level off to a value largely set by the Planck selection function,
as shown in the central panel by the green curve. The discontinuity at $\log M /M_{\odot}\sim 14.3$ 
in the central panel is
due to the change of mean redshift, and therefore of Planck limiting mass, of
clusters of that richness-based mass. At high masses, 
$E(\log M_{Planck}|\log M_{richness})$ converges to $M_{200}=1.3 M_{500}$
(blue line). In the right--hand panel of Fig.~B.1, the flattening above is absent
because our catalog of richness-based masses is limited to a mass lower than the Planck
limit, and therefore the relation is a 
straight line in the right--hand panel. We predict a bending of this line
at $\log M_{200} /M_{\odot}\sim 14$ in the future when a 
comparison sample of less massive clusters will be available, precisely
as the Planck subsample does at $\log M_{500} /M_{\odot}\sim 14.4$
when compared to our richness-based masses.
This line differs from $M_{200}=1.3 M_{500}$ because this relation holds
for true values, while we are using {\it observed} values\footnote{In a
scattered relation, $E(x|y)$ differs from $E(y|x)$, and both differ from the ``true"
relation between $x$ and $y$, see Fig.~1 in Andreon \& Hurn 2010 for an
example.}. 

We now consider the true data, i.e., the dots in central and right--hand
panels of Fig.~B.1, that are mean values in bins of 8 clusters in the catalog
of the 107 clusters in the overlap of our catalog and the Planck catalog, 
starting the binning from the most massive cluster. At low masses,
they follow the expected behavior, i.e., the green curves computed
using the simulated data above, including
the point with the lowest mass  in the central panel, the latter due 
to the change
of mean redshift. 
At high masses, the right--hand and central panels show  an
indication of an underestimation of Planck masses 
(lower than expectations, i.e., below the green line), in agreement with the
mass--dependent bias suggested by
von der Linden et al. (2014), Andreon (2015), and Sereno et al. (2015).

A naive inspection of the three panels would have lead to the
wrong conclusions. In fact,
masses deviate from the naive expectation
($M_{200}=1.3 M_{500}$, the blue line, plotted in two of the panels)
at low masses in all panels; however,
this expectation is inappropriate for these data.
An orthogonal or bisector fit would have incorrectly returned an agreement at high
masses using the individual data points (left--hand panel), a mass tilt at low masses
if $E(\log M_{Planck}|\log M_{richness})$ were used (central panel), and an
agreement at high masses if $E(\log M_{richness}|\log M_{Planck})$ were used.
To summarize, fitting the trend of two estimates of mass, or
determining a mass bias, it is not just a matter of fitting the data cloud 
with an off-the-shelf (linear) fitter, unless the selection function
is ignorable and the scatter/errors of the masses are negligible. 

The use of the cosmological sample (61 clusters) does not alter our conclusions above,
i.e., there is agreement with Planck masses at low masses only and
the data trend may well be different from naive expectations.

We emphasize that for Planck clusters
the expected trend can be computed because both types of masses are
homogeneously measured  and because of the
controlled nature of the subsample being studied, having a known selection function. 
Such predictions, necessary to
discriminate real trends from selection effects, are not possible with
samples with unknown selection functions or with non-homogeneously
measured masses, such as the current sample of clusters
with weak--lensing masses.

\section{List of blended clusters}

\begin{table*}
\caption{Clusters blended on the $r_{200}$ spatial scale.}
\footnotesize
\begin{tabular}{l l l l l l }
\hline
R.A. & Dec. & Piffaretti et al. 2011 ID & Planck 2015 ID & Abell 1957 ID & NED ID  \\
 \multicolumn{2}{c}{J2000} & & &  \\
\hline
137.3130 & 11.0072 &  MCXCJ0909.1+1059 (2.0) & PSZ2G218.81+35.51 (2.0)   &   ABELL750 (2.5) & MS 0906.5+1110 (1.8,48) 	\\
158.0309 & 40.2137 &                         &  			 &  ABELL1035 (0.3) & ABELL1035 (0.3,63)  \\  
170.3957 & 48.0605 &  MCXCJ1121.5+4803 (0.2) &  			 &  ABELL1227 (2.1) & ABELL1227 (2.1,34)  \\
172.5019 & 20.4423 &  MCXCJ1130.0+2028 (2.4) &  			 &  ABELL1278 (3.7) & ABELL1278 (2.4,25)  \\ 
189.1022 & 16.5624 &  MCXCJ1236.4+1631 (2.0) &  			 &  ABELL1569 (2.3) & ABELL1569 (2.3,57)  \\ 
205.4807 & 26.3703 &  MCXCJ1341.8+2622 (0.4) & PSZ2G031.93+78.71 (0.6)   &  ABELL1775 (0.3) & ABELL1775 (0.4,161) \\  
206.5196 & 54.0439 &                         &  			 &  ABELL1790 (3.1) & ABELL1790 (3.1,16)  \\ 
209.7823 & 27.9720 &  MCXCJ1359.2+2758 (2.4) & PSZ2G040.03+74.95 (2.1)   &  ABELL1831 (1.3) & ABELL1831 (2.4,109) \\  
227.8604 & 18.0723 &                         &  			 &  ABELL2036 (1.0) & ABELL2036 (1.0,33)  \\ 
240.3538 & 53.9057 &  MCXCJ1601.3+5354 (0.4) &  			 &  ABELL2149 (3.7) & ABELL2149 (3.7,58)  \\ 
\hline      
\end{tabular}
\hfill\break
The table lists cluster coordinates (J2000), followed by the id of clusters
from a few bibliographic sources, with angular offsets reported in parentheses. The
last column also lists the number of references to the cluster in Mai 2015, as given by NED. 
\end{table*}

\end{document}